%% file: main.tex
\documentclass[aps, pra,twocolumn]{revtex4-2} 

\usepackage[utf8]{inputenc}
\usepackage{graphicx}
\usepackage{dcolumn}
\usepackage{bm}
\usepackage{colortbl}  
\usepackage{diagbox}   
\usepackage[table]{xcolor}
\usepackage{amsthm}
\usepackage{amsmath}
\usepackage{amssymb}
\usepackage{mathrsfs}
\usepackage{comment}
\usepackage{color}
\usepackage[breaklinks=true,colorlinks,allcolors=blue]{hyperref}
\usepackage{braket}
\usepackage{stackengine}
\usepackage{scalerel}
\usepackage{dsfont}
\usepackage{float}
\usepackage{tikz}
\usepackage{tabularx}
\usepackage{booktabs}   

\usepackage{multirow}
\usepackage{array}

\usepackage{ragged2e}
\usepackage{etoolbox}

\usepackage[paperwidth=210mm,paperheight=297mm,hmargin=1.85cm,vmargin=2.5cm]{geometry}

\begin{document}

\preprint{APS/123-QED} 

\title{Quantum Energetic Advantage before Computational Advantage in Boson Sampling}

\author{Ariane Soret$^{\dagger}$, Nessim Dridi$^{\dagger, \ddagger}$, Stephen C. Wein$^{\dagger}$, Valérian Giesz$^{\dagger}$, Shane  Mansfield$^{\dagger}$, Pierre-Emmanuel Emeriau$^{\dagger}$}
\email{pe.emeriau@quandela.com} 
\affiliation{$^\dagger$ Quandela SAS, 7 Rue Léonard de Vinci, 91300 Massy, France \\
$ ^\ddagger$Eidgenössische Technische Hochschule Zürich, Rämistrasse 101, 8092 Zürich, Switzerland}

\date{\today}

\begin{abstract}
Understanding the energetic efficiency of quantum computers is essential for assessing their scalability and for determining whether quantum technologies can outperform classical computation beyond runtime alone. In this work, we analyze the energy cost required to solve the Boson Sampling problem, a paradigmatic task for quantum advantage, using a realistic photonic quantum computing architecture. Using the Metric–Noise–Resource methodology, we establish a quantitative connection between experimental control parameters, dominant noise processes, and energetic resources through a performance metric tailored to Boson Sampling. We estimate the energy cost per sample and identify operating regimes that optimize energetic efficiency. By comparing the energy consumption of quantum and state-of-the-art classical implementations, we demonstrate the existence of a quantum energetic advantage -- defined as a lower energy cost per sample compared to the best-known classical implementation -- that emerges before the onset of computational advantage, even in regimes where classical algorithms remain faster. Finally, we propose an experimentally feasible Boson Sampling architecture, including a complete noise and loss budget, that enables a near-term observation of quantum energetic advantage.
\end{abstract}

\maketitle

\input{sections/0-introduction}

\input{sections/1-MNR}

\input{sections/2-Boson-sampling}

\input{sections/3-Energetic-advantage}

\input{sections/4-Hardware}

\input{sections/5-sycamore}

\input{sections/6-conclusion}

\bibliography{bib}
\clearpage

\input{sections/7-appendix}

\end{document}

%% file: sections/0-introduction.tex
\section{Introduction}
\label{sec:intro}
The rapid development of quantum computing in the past decades has been primarily driven by the prospect of \textit{computational advantage}, that is, their ability to solve problems faster than classical computers. 
Among the early milestones, \textit{Boson Sampling} \cite{aaronson2011computational} emerged as a paradigmatic example demonstrating that a simple, non-universal photonic processor could perform a task believed to be intractable for classical computers. Since its introduction, Boson Sampling has played a dual role as both a theoretical benchmark \cite{clifford2018classical, renema2018classical, go2025bs, oszmaniec2025complexitytheoreticfoundationsbosonsamplinglinear} and an experimental testbed \cite{wu2018benchmark, wang2019bs-20-photons} for near-term quantum speedups.

Beyond runtime considerations, however, an increasingly important question concerns the energetic efficiency of quantum computers. As quantum devices scale up, their viability will depend not only on speed, but also on whether they can perform computational tasks using fewer physical resources (notably energy) than their classical counterparts. This perspective motivates the notion of \textit{quantum energetic advantage} \cite{Auff_ves_2022}, which asks whether quantum hardware can outperform classical systems in energy consumption for a given task, independently of whether it does so in time. Crucially, energetic advantage need not coincide with computational advantage, and understanding their relationship is essential for evaluating the practical impact of quantum technologies.

Recent work has begun to formalize energetic considerations in quantum information processing, drawing connections between quantum thermodynamics, control theory, and resource optimization \cite{fellous-asiani2021limitations,huard2022qubitgate,buffoni2022thirdlaw,fellous2023optimizing,singh2023proof,volpatoEstimatingElectricalEnergy2024,bilokur2024thermodynamiclimitationsfaulttolerantquantum,meier2025energyconsumptionadvantagequantumcomputation,yehia2025energeticemergingquantum, upreti2025boundingcomputationalpowerbosonic, zhao2025learningerasequantumstates, thompson2025energeticadvantagesquantumagents, ishida2025quantumcomputerbasedverificationquantum, badhani2025thermodynamicworkcapacityquantum}. These studies have explored energy costs at the level of gates, qubits, and abstract computational models. However, analyses that connect full-stack, hardware-level energy consumption to the performance of concrete quantum algorithms remain scarce. In particular, it is still largely unexplored for which tasks quantum energetic advantages can arise in realistic architectures before the onset of computational advantage, when classical algorithms remain faster.

In this work, we address this question using Boson Sampling as a case study. While Boson Sampling is historically associated with demonstrations of computational hardness, it is also particularly well suited for energetic analysis: sampling is native to the quantum hardware, incurs minimal algorithmic overhead, and can be repeated at high rates, whereas classical simulation requires increasingly expensive numerical procedures as the effective problem size grows. These features make Boson Sampling an ideal platform for investigating whether energetic advantage can precede computational advantage in practice.

To carry out this analysis, we adopt the Metric–Noise–Resource (MNR) framework \cite{fellous2023optimizing}, which provides a systematic methodology to relate experimental control parameters, dominant noise mechanisms, and resource consumption to a task-specific performance metrics. This approach allows us to go beyond simplified energy-per-operation models by capturing the interplay between hardware constraints, noise mitigation strategies, and algorithmic performance at the system level. We specialize the MNR framework to photonic implementations of Boson Sampling based on single-photon sources, linear optical interferometers, and single-photon detectors, and we define a performance metric that reflects the complexity of the best-known classical simulation algorithms in the presence of loss and partial distinguishability.

Using this framework, we quantitatively estimate the energy cost per sample of a photonic Boson Sampling experiment and identify operating regimes that optimize energetic efficiency. By comparing these results with state-of-the-art classical simulation strategies running on energy-efficient supercomputing hardware, we demonstrate the existence of a quantum energetic advantage that emerges before the threshold for quantum computational advantage is reached. In this regime, photonic quantum devices consume less energy per sample than classical computers, even though classical algorithms remain faster.

Finally, we translate these findings into experimentally relevant requirements by providing a detailed noise and loss budget for a near-term Boson Sampling architecture. This analysis identifies a realistic parameter regime in which a quantum energetic advantage can be observed experimentally, thereby establishing energetic efficiency as an independent and experimentally accessible benchmark for quantum advantage.

\begin{figure*}[htbp]
    \centering
    \includegraphics[width=0.9\textwidth]{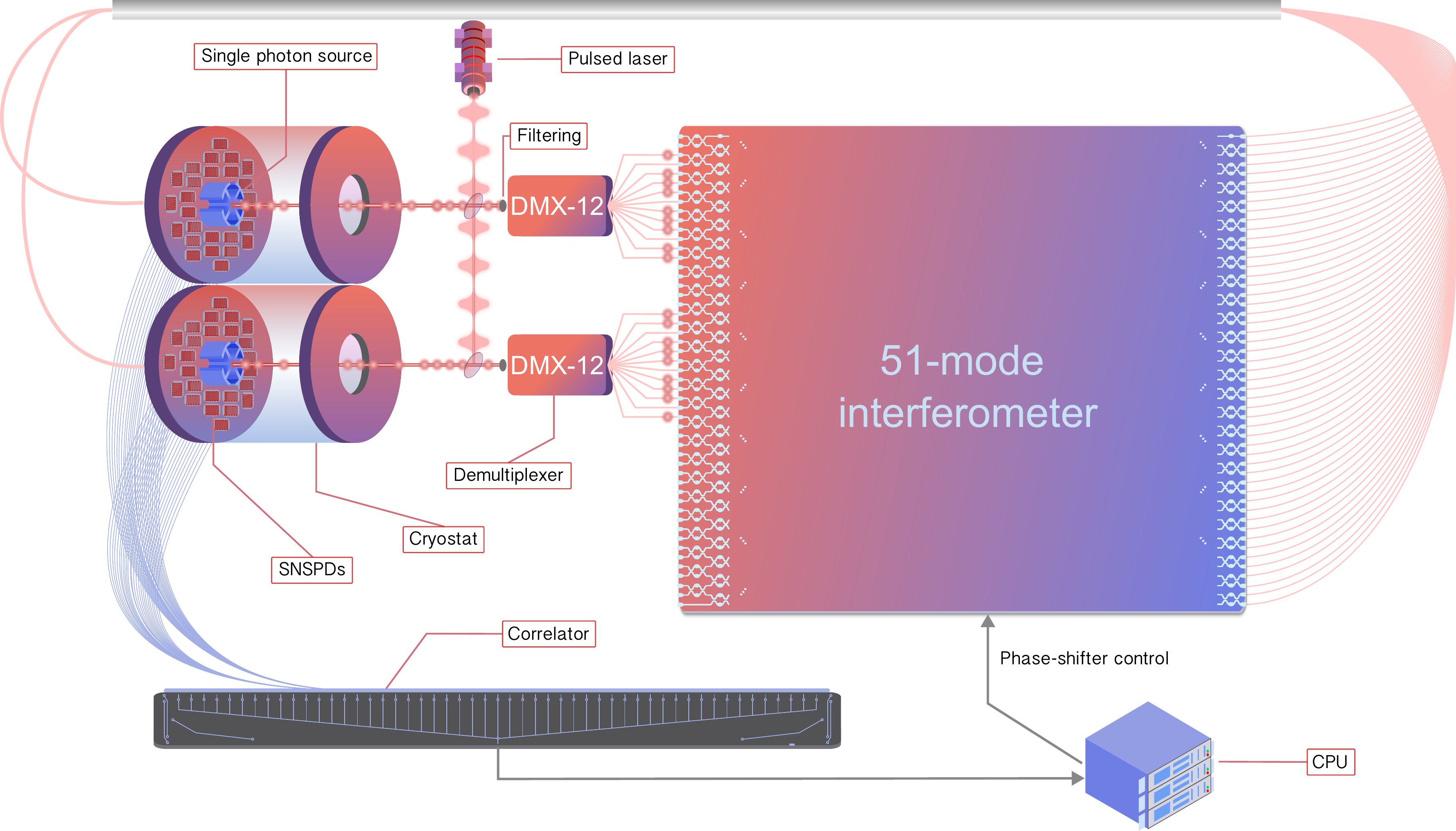}
    \caption{
    \textit{Proposed photonic Boson Sampling architecture for observing a quantum energetic advantage.}
    Two quantum-dot single-photon sources operated at 3 K emit trains of single photons at a 1 GHz repetition rate. Active demultiplexers and fiber delays are used to inject up to 24 photons simultaneously into a 51-mode linear optical interferometer. The photons are detected using superconducting nanowire single-photon detectors (SNSPDs), with detection events processed by classical electronics. The architecture enables a full-stack estimation of the energy cost per generated sample, including cryogenic cooling, photon generation, interferometric processing, and detection. The end-to-end transmission and noise budget required to observe energetic advantage are detailed in Table\ref{table:source-eff-target}.
    }
    \label{fig:circuit}
\end{figure*}

\medskip

%% file: sections/1-MNR.tex
\section{Metric-Noise-Resource}
\label{sec:mnr}
The MNR framework is designed to quantify and optimize the resource consumption of quantum computers \cite{fellous2023optimizing}. It provides a systematic approach to relate a performance metric ($\mathcal{M}$) of a given quantum computation to the underlying noise processes ($\mathcal{N}$) and the resources ($\mathcal{R}$) required to control and compensate for the noise. The key elements of MNR are the following:

\begin{enumerate}

\item \textbf{Control Parameters}: The set of variables $\vec{C} = (C_1, C_2, \dots, C_d)$ that can be adjusted to influence the performance and resource consumption of the quantum computer. Control parameters can be hardware-related, such as the qubit temperature or control line attenuation, and software-related, such as circuit depth, gate sequences, or error-correcting code parameters.

\item \textbf{Metric}: The performance metric $\mathcal{M}$ quantifies the quality or success probability of the quantum computation, higher values of $\mathcal{M}$ indicating a better performance. For example: the gate fidelity, error rates or success probabilities.

\item \textbf{Noise}: The noise $\mathcal{N}$ denotes the physical processes that decrease the performance of quantum computations. Typically, the noise is modeled using master equations or noise channels that depend on the control parameters. Noise mitigation strategies, such as cooling qubits to lower temperatures or implementing error correction codes, can reduce the noise but often come at the cost of increased resource consumption.

\item \textbf{Resource}: The resource consumption $\mathcal{R}$ encompasses the physical and computational resources required to execute the quantum computation. This may include energy consumption, gate operations, and classical control overhead.

\end{enumerate}
The metric, noise, and resources are expressed in terms of the control parameters. The idea of the MNR methodology is to identify the set of control parameters that minimizes the resources while aiming at a fixed target metric $M_0$,
\begin{align*}
\min_{\vec{C}} \mathcal{R}(\vec{C}) \quad \text{s.t.} \quad \mathcal{M}(\vec{C}) = M_0 \, .
\end{align*}
Alternatively, one can consider maximizing the efficiency, defined as the ratio of the performance metric to the resource consumption, under the same constraint,
\begin{align}
\max_{\vec{C}} \frac{\mathcal{M}(\vec{C})}{\mathcal{R}(\vec{C})} \quad \text{s.t.} \quad \mathcal{M}(\vec{C}) = M_0 \, .
\label{eq:efficiency_optimization}
\end{align}
Examining this constrained optimization problem allows the exploration of trade-offs between performance and resource costs, and the identification of optimal operational regimes.
\medskip

%% file: sections/2-Boson-sampling.tex
\section{MNR for Boson Sampling}
\label{sec:mnrBS}
Boson Sampling is a quantum computing problem that was first proposed by Aaronson and Arkhipov as a practical way to achieve a quantum computational advantage \cite{aaronson2011computational}, and has since been demonstrated experimentally \cite{madsen2022bs-demonstration,wang2019bs-20-photons}. The Boson Sampling problem is defined as follows: given an input consisting of \( n \) single photons and a Haar random interferometer described by an \( m \times m \) unitary matrix \( U \) ($m \gg n^2$), sample from the output distribution of the photons after they have passed through the interferometer.
The unitary matrix \( U \) represents the linear optical network through which the photons propagate. 
Let \( S = \{s_1, s_2, \ldots, s_m\} \) be the occupation-number tuple representing a possible input configuration of the photons.
The \( s_i \in \{0, \ldots, n\}\) represent the number of photons in mode \(i\) so that \(\sum_{i=1}^m s_i = n\). 
The task is to sample from the distribution over output configurations \( V = \{v_1, v_2, \ldots, v_m\} \), where \( v_i \in \{0, \ldots, n\} \). 
The probability of observing a specific output configuration \( V \) given an input configuration \( S \) is related to the permanent of a submatrix, \( U_V^S \), of \( U \), obtained by repeating $s_i$ times the $i^\text{th}$ column of $U$, and then repeating $v_j$ times the $j^\text{th}$ row of the resulting matrix. 
Then,
\begin{align*}
P(V|S) = \frac{|\text{Per}(U_V^S)|^2}{s_1! s_2! \ldots s_m! v_1! v_2! \ldots v_m!},
\end{align*}
where \(\text{Per}(U_V^S)\) denotes the permanent of \( U_V^S \). Calculating the permanent of a matrix is a \#P-hard problem in general \cite{aaronson2011computational}, making exact Boson Sampling intractable for classical computers as the size of the system grows \footnote{Specifically, Aaronson and Arkhipov demonstrated that if an efficient classical algorithm for exact Boson Sampling existed, it would imply that \( P^{\#P} = BPP^{NP} \), leading to a collapse of the polynomial hierarchy in computational complexity theory \cite{aaronson2011computational}.}. The proof also extends to the approximate case, which is the realistic scenario for a quantum device.  
Therefore, approximate Boson Sampling is believed to be efficiently solvable only by quantum computers.

Let us now apply MNR to Boson Sampling. 
We consider a Boson Sampling setup consisting of single-photon sources followed by demultiplexers, an \(m\times m\) interferometer and \(m\) single-photon detectors.
The sources and detectors are placed inside the same cryostat at temperature $T$. 
Let $n$ be the number of input photons. 

\textbf{Control parameters:}
The control parameters are the number of input photons, the number of modes, and the temperature of the cryostat. In this work, we impose that \(m = \lceil 2.1n\rceil \), implying that we are \textit{not} in the collision-free regime (originally characterized by $m\gg n^2$ \cite{aaronson2011computational}). The collision-free regime --- in which the probability of two photons reaching the same output mode is negligible --- has long been considered necessary for demonstrations of computational advantage. However, it has recently been proven that one can relax this assumption, and that a linear scaling $m\geq 2.1n$ is sufficient \cite{oszmaniec2025complexitytheoreticfoundationsbosonsamplinglinear}, hence our choice of $m=\lceil 2.1n\rceil$. We can then identify our control parameters as \(\vec{C} = (T, n)\).
    
\textbf{Resource:} The hardware of reference for this work is a photonic quantum computer using a quantum-dot based single-photon source \cite{senellart2021bright} with superconducting nanowire single-photon detectors (SNSPD) \cite{maring2024versatile}. 
Currently, cryogenic cooling needed to operate the single-photon source and SNSPDs below 3\,K takes up the largest share of the overall power consumption of these technologies \cite{maring2024versatile}.
Since heat dissipated by the detectors and the single-photon source (dynamic load) is negligible, the dominant cryogenic cost comes from static heat leaks, meaning that it does not depend on whether the SNSPDs are on or off. Since there is a maximum number, $s_\text{max}$, of SNSPDs which we can store in a cryostat, and given that we need one SNSPD per mode, the total cryogenic power is related to $m$ through  
\begin{equation}
    P_{\text{cryo}}(T,m) = \frac{P_0}{\eta_c(T)}\left(1+\left\lfloor\frac{m}{s_\text{max}} \right\rfloor \right ) \, ,
    \label{eq:p-cryo}
\end{equation}
where $P_0/\eta_c$ is the power of a single cryostat, with $P_0$ the cooling power and $\eta_c=xT/(T_\text{ext}-T)$ the cryostat efficiency, expressed as a percentage $x$ of the Carnot efficiency, with $T_\text{ext}=300$\, K the external temperature. Finally, $1+\lfloor m/s_\text{max}\rfloor$ counts the number of cryostats required. 
In addition to cryogenic power, electric power is consumed for the laser, electronic monitoring and classical overhead, which we estimate from experiments as a fixed cost $P_\text{fix}$. 
The resource cost \(\mathcal{R}\) is therefore 
\begin{equation}
    P_{\text{tot}}(T,n) = P_\text{fix} + P_{\text{cryo}}(T,\lceil 2.1n\rceil) \, ,
    \label{eq:p-tot}
\end{equation}
where we used the relation $m=\lceil 2.1n\rceil$.

The cryogenic model captures the dominant energy costs of current and near-term photonic quantum computing platforms in which static heat loads associated with cryostat operation largely dominate over dynamic dissipation. More integrated architectures, alternative cooling strategies, or higher detector densities per cryostat would modify the quantitative scaling of the cryogenic cost, but would not alter the qualitative conclusions of our analysis. In particular, reducing the cryogenic overhead per mode would only further improve the energetic performance.

\textbf{Noise:} The experimental implementation of Boson Sampling on a photonic quantum computer is mainly subject to two sources of noise: photon loss and photon distinguishability. 

For now, we assume that a fixed number $l$ of photons is lost. We will relate $l$ to the characteristics of the hardware in the paragraph on hardware requirements. 
 
Photon distinguishability, on the other hand, stems from the interaction between the source and the environment, and from imperfections of the source, which induce changes in internal photon degrees of freedom (e.g., polarization). For semiconductor single-photon quantum dot sources, decoherence is mainly induced by charge and spin noise in the semiconductor \cite{warburton2013spinchargenoise}. However, such noises can be reduced using ultra-clean materials (for charge noise) and dynamic decoupling schemes (for spin noise). In addition, acoustic phonons, which propagate in the semiconductor, induce two dephasing processes: first, a fast decay in the zero-phonon line (ZPL), second, a dephasing of the ZPL itself. Typically, quantum dots emit about 90\% of the photons in the ZPL \cite{grange2017reducing}. When the quantum dot is coupled to a high quality factor cavity, the coupling enhances the radiative emission rate through the Purcell effect, and redirects the phonon sideband (PSB) photons into the ZPL \cite{grange2017reducing}. As a result, the fraction of emission in the ZPL, \(\eta_{\text{ZPL}}\), increases significantly, thereby enhancing the overall indistinguishability. The indistinguishability of the ZPL, $I_\text{ZPL}$, then becomes a good approximation for the indistinguishability $I$ of the photons emitted by a single-photon source. We shall use this approximation in this work.
\(I_\text{ZPL}\) can be written as \cite{grange2017reducing, grange2017reducingsupp},
\begin{align}
\label{eq:ZPL}
I_\text{ZPL}(T) = \frac{\gamma(T)}{\Gamma(T) } \, ,
\end{align}
where $\Gamma(T)\equiv \gamma(T) + \gamma^*(T)$, with $\gamma(T)$ the natural linewidth -- the emission rate in the cavity -- and \(\gamma^*(T)\) the additional broadening of the ZPL due to pure dephasing. $\gamma^*(T)$ increases with temperature, while $\gamma(T)$ can be enhanced through the Purcell effect. We refer to the Appendix~\ref{app:dist} for analytical expressions. The key point is the fact that the indistinguishability depends on the temperature. When considering multiple sources, additional sources of noise have to be taken into account: the spectral detuning between the sources, and the different temporal shapes of the pulses. Moreover, the sources can have slow frequency fluctuations (over timescales larger than $1/\gamma$), implying that the detuning should be treated as a random variable. The average indistinguishability is then obtained after averaging over the detuning. In this work, we assume that the sources are nearly identical and that the noise due to pure dephasing dominates, which allows us to use the $I_\text{ZPL}$ of a single source as an upper bound (see Appendix~\ref{app:dist} for details). Then, we account for the slow spectral fluctuations by removing 5\% from the indistinguishability, in accordance with the state-of-the-art values obtained in \cite{zhai2022interferenceremoteqd},
\begin{align}
\label{eq:ZPL-2qd}
I(T) \approx  I_\text{ZPL}(T)-0.05\, .
\end{align}

\textbf{Metric:} A natural metric is one that reflects the cost of the best classical algorithm to perform Boson Sampling.
A classical algorithm for simulating Boson Sampling with partially distinguishable photons was proposed by \cite{renema2018efficient}. This algorithm was further developed to also take into account loss \cite{renema2018classical}. Both these algorithms have been proposed for the collision-free regime, $m \gg n^2$. However they can be extended to the saturated regime, $m=\Theta(n)$. As shown in \cite{oszmaniec2025complexitytheoreticfoundationsbosonsamplinglinear}, regardless of the regime (collision-free, saturated, or any other), the complexity of Boson Sampling depends on the number of detectors triggered, $c_S$, rather than on the number of input photons $n$. The outcomes of a Boson Sampling experiment with $n$ photons and $m$ modes (without losses) typically trigger 
\begin{equation}
    c_S(n,m)=\frac{m/n}{m/n+1}n
    \label{eq:cs}
\end{equation}
detectors. More precisely, the probability to observe an outcome which triggers $c_S$ detectors follows a large deviation principle, $\text{Pr}[|c_S-\frac{m/n}{m/n+1+1/n}n|\geq tn]\leq 2e^{-2t^2n}$ for any $t\geq 0$. Notice that Eq.~\eqref{eq:cs} can be generalized to the case of Boson Sampling with losses, i.e. with $n-l$ photons, and becomes 
\begin{equation}
    c_S^l(n,m)=\frac{m/[n-l]}{m/[n-l]+1}[n-l] \, .
    \label{eq:csl}
\end{equation}
Remarkably Boson Sampling is \# P-hard both in the collision-free regime (which was known) and in the saturated regime, with the size of the problem being defined by $c_S$. We may then adapt the algorithm of \cite{renema2018classical} to a problem of size $c_S$ rather than $n$. A crucial aspect of the work of \cite{renema2018classical} involves the quantity \( k \), which characterizes the ability to efficiently solve Boson Sampling classically for a fixed number of lost photons $l$. In the collision-free regime, the value of $k$ depends on $n$ and $T$:
\begin{align}
    k(T,n) = \frac{\ln \left\{ \frac{\varepsilon \delta}{2} \left[ 1 - I(T)^2\frac{ n-l}{n} \right] \right\}}{\ln \left[ I(T)^2 \frac{n-l}{n} \right]} - 1
    \label{eq:k}
\end{align}
where $I$ is defined in Eq.~\eqref{eq:ZPL-2qd}, $\varepsilon$ represents the error tolerance in the classical simulation and $\delta$ is the probability of failure of the algorithm. We fix $\varepsilon \delta = 0.001$, following a choice discussed in \cite{renema2018classical}. The number $k$ can be interpreted as follows: a noisy $n$-photon Boson Sampling task is equivalent to another Boson Sampling task, with $k$ perfect photons, and $n-k$ completely distinguishable particles that behave like classical particles.  
In other words, the hardness of simulating the imperfect case is limited to simulating less bosons in the perfect case. Let us now adapt this reasoning to the saturated regime: consider a Boson Sampling problem with $n$ noisy photons, with indistinguishability $I$, in a circuit such that $m=\Theta(n)$ with $l$ lost photons. In the perfectly indistinguishable and lossless case, the size of the Boson Sampling problem would be $c_S(n,m)$ in Eq.~\eqref{eq:cs}. With losses, the size of the problem is $c_S^l(n,m)$ in Eq.~\eqref{eq:csl}. Adding now partial distinguishability, we deduce from Eq.~\eqref{eq:k} that our initial Boson Sampling task is equivalent to a Boson Sampling task with 
\begin{equation}
    k^\text{eff}(T,n,m) = \frac{\ln \left\{ \frac{\varepsilon \delta}{2} \left[ 1 - I(T)^2\frac{c_S^l(n,m)}{c_S(n,m)} \right] \right\}}{\ln \left[ I(T)^2 \frac{c_S^l(n,m)}{c_S(n,m)} \right]} - 1
    \label{eq:keff}
\end{equation}
perfectly indistinguishable photons and $c_S^l(n,m)-k^\text{eff}(T,n,m)$ fully distinguishable particles. This concludes the generalization. 

We are now in position to define the metric. Since we impose $m=\lceil 2.1n\rceil$, and assume a fixed $l$, the functions $c_S, c_S^l$ and $k^\text{eff}$ depend on $T,n$ only. We only allow physical solutions for the optimization problem, requiring that $n-l \geq k^\text{eff}$. This leads to the definition of our performance metric:
\begin{align}
    \mathcal{M}(T,n) = \min[n-l(n), k^\text{eff}(T,n)] \, .
    \label{eq:metric}
\end{align}
Given these definitions, we can anticipate the following trade-off: assuming the power consumption is held constant, increasing the number of photons \( n \), and hence of modes \( m \), in the Boson Sampling experiment, requires operating the cryostats at a higher temperature \( T \) whenever an additional cryostat is required. 
However, increasing \(T\) leads to higher phonon interactions, which reduces the photon indistinguishability \(I\), hence reducing \(\mathcal{M}\). Consequently, we expect an interplay between the system size (\(m \)), operational temperature ($T$), and photon indistinguishability (\(I\)).

\medskip

%% file: sections/3-Energetic-advantage.tex
\section{Quantum energetic advantage for Boson Sampling}
\subsection{Classical vs.\ quantum comparison}
Before comparing quantum and classical energy costs, we emphasize that energetic advantage constitutes an independent notion of quantum advantage. A quantum device may consume less energy per sample than a classical computer even when classical algorithms remain faster in runtime. Conversely, computational advantage does not imply energetic advantage, since energy consumption depends on hardware architecture, control resources, and the strategies required to mitigate noise.

We now reach the core of the paper: the comparison of the energy cost per sample of a quantum computer performing Boson Sampling, \(E_{\text{sample}}^Q\), to that of a classical supercomputer executing a classical algorithm for the same problem, \(E_{\text{sample}}^C\). The method is as follows: first, we perform optimizations assuming $l$ lost photons for $l\in[1,25]$. 
Then, we take into account the fluctuations in the number of transmitted photons, and we compare the \textit{average} energy per sample, for a quantum and a classical computer, to produce a set of samples, 
where the number of transmitted photons fluctuates according to a binomial distribution.  

\begin{figure*}[ht!]
    \centering
  \hspace*{-1.8cm}  \includegraphics[width=1.2\linewidth]{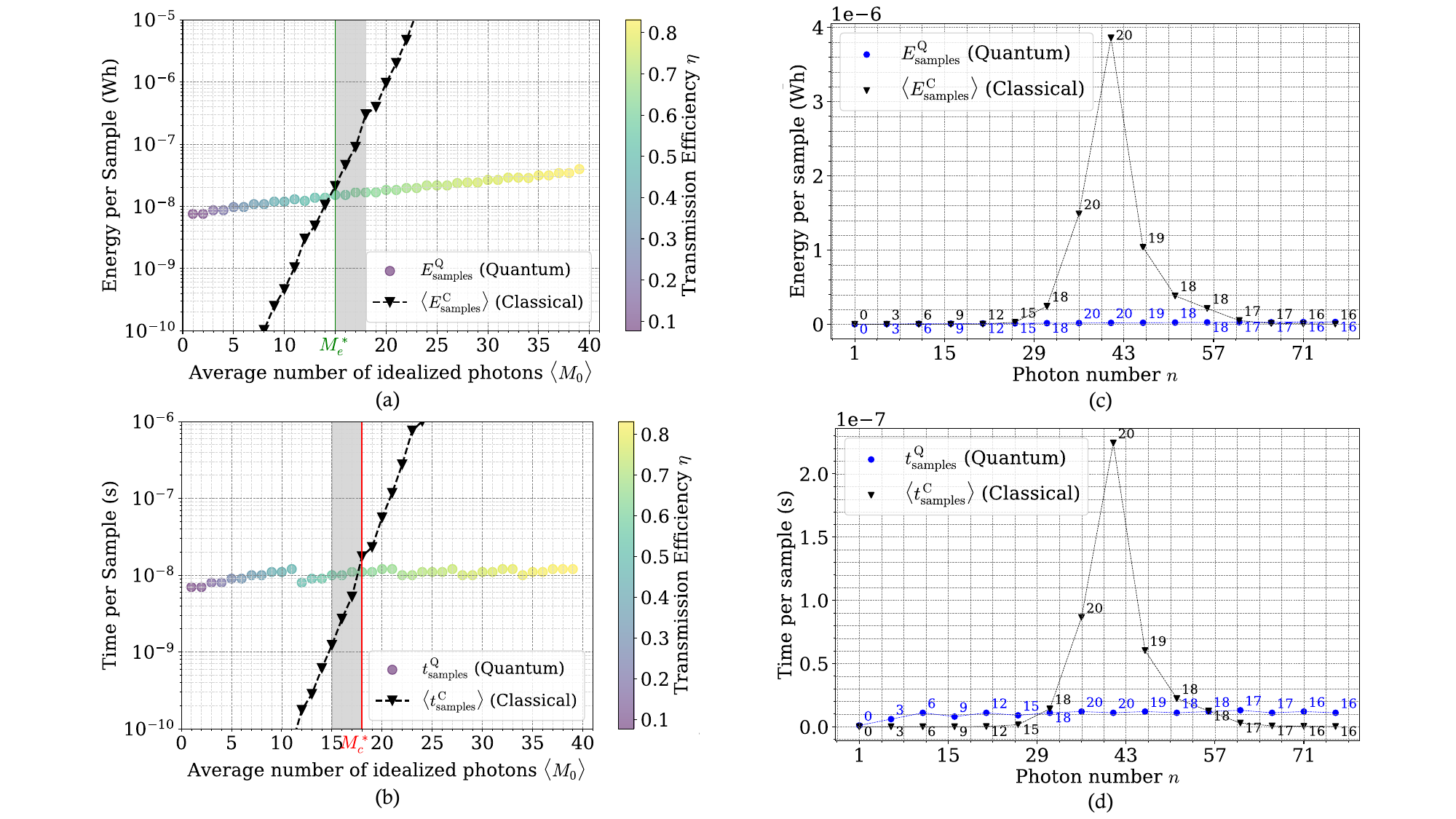}
  \caption{
    \textit{Quantum energetic advantage preceding computational advantage in Boson Sampling.}
    (a) Energy cost per sample for the photonic quantum device ($E^\text{Q}_\text{samples}$) and for classical simulation ($\langle E^\text{C}_\text{samples}\rangle$), shown as a function of the average performance metric $\langle M_0\rangle$. Quantum data points are colored according to the end-to-end transmission efficiency.
    (b) Corresponding time per sample for quantum ($t^\text{Q}_\text{samples}$  samples) and classical ($\langle t^\text{C}_\text{samples}\rangle$) implementations. The shaded region highlights the regime in which the quantum device consumes less energy per sample than the classical simulation, despite remaining slower in runtime.
    (c,d) Energy and time per sample as functions of the number of input photons for a realistic hardware configuration (Fig.~\ref{fig:circuit}) with fixed per-component losses, illustrating the degradation of the metric at large system sizes when transmission is not improved.
  }
        \label{fig:results}
\end{figure*}

\subsubsection{Classical Energetic Cost per Sample}

In order to obtain \(E_{\text{sample}}^C\), we consider the fast and exact classical algorithm for Boson Sampling proposed by Clifford and Clifford, which has a time complexity of \(\mathcal{O}(n 2^n+mn^2) \) \cite{clifford2018classical}. Our objective is to determine a lower bound on the number of floating-point operations required by this algorithm to generate a single sample. We identify the two most computationally expensive steps in the algorithm as:
\begin{itemize}
    \item \textbf{Step 1:} Computing the coefficients of the Laplace expansion of the permanent of a \(j \times j\) matrix, which requires \(j2^j\) floating-point operations.
    \item \textbf{Step 2:} Obtaining an array of length \(m\) by summing \(j\) terms \(m\) times. This requires \(mj\) floating-point operations.
\end{itemize}
These two steps are repeated for each \(j \in \{2,...,n\}\). The total number of floating-point operations required for executing the algorithm is greater than the sum of the contributions from these two steps only. Therefore, a lower bound $L(m,n)$ on the number of floating-point operations is
\begin{equation}
L(m, n) \equiv \sum_{j=2}^{n} \left( j2^j + mj \right) \, .
\end{equation}
The energy required to perform $L(m,n)$ operations depends on the energy efficiency of the classical hardware. For a given efficiency $\eta_e$ in Flops per watt, the energy consumption of the classical algorithm per sample is then $E_{\text{sample}}^C(m, n) = \frac{L(m, n)}{\eta_e}$. 
Since we are working with partially distinguishable photons, the computational complexity of the classical simulation is represented by the metric \(\mathcal{M}\) rather than the number of input photons \(n\). Therefore, we should replace \(n\) with \(\mathcal{M}\), which yields an expression in terms of the control parameters $T,n$,
\begin{equation}
E_{\text{sample}}^C(T,n) = \frac{L[\lceil 2.1 n\rceil,\mathcal{M}(T,n)]}{\eta_e} \, .
\label{eq:classical-energy-per-sample-2}
\end{equation}

Note that we deliberately adopt conservative assumptions in favor of the classical simulation. First, we estimate the classical energy cost using a lower bound on the number of floating-point operations required by the simulation algorithm. Second, we benchmark against one of the most energy-efficient supercomputing platforms currently available. Third, we neglect additional sources of classical overhead such as memory access, communication costs, and scheduling inefficiencies, which are known to contribute significantly to the energy consumption of large-scale simulations. As a result, the classical energy costs reported here should be regarded as optimistic lower bounds.

\subsubsection{Quantum Energetic Cost per Sample}

The quantum energy cost per sample is the product of the total power consumption Eq.~\eqref{eq:p-tot} and the time required to produce a sample, i.e. the time required to emit $n$ photons. The latter is obtained from the sample generation rate $r_{\text{sample}}$. Assuming that we have one source per cryostat, the rate will be limited by the maximum number of photons that has to be sent by one source, which is, given $n$ and $m$: $\overline{n} = \lceil \frac{n}{1+\left\lfloor\frac{m}{s_\text{max}}\right\rfloor} \rceil$. The rate is then given by:
\begin{equation}
r_{\text{sample}}(n) = \frac{r_{\text{SPS}}}{\overline{n}},
\label{eq:q-rate}
\end{equation}
where $r_{\text{SPS}}$ is the single-photon generation rate.   
The energy per sample $E_{\text{sample}}^Q$ is then 
\begin{equation}
E_{\text{sample}}^Q(T,n) = \frac{P_{\text{tot}}(T,n)}{r_{\text{sample}}(n)} .
\label{eq:quantum-energy-per-sample}
\end{equation}

\subsubsection{Quantum Energetic Advantage in Boson Sampling}

In order to make a quantitative analysis, we now need to make some hardware assumptions. For the classical hardware, we consider the most energy-efficient system ranked in the Green500 list, the JEDI module of the exascale supercomputer JUPITER \cite{green500_2024}. This system has an energy efficiency \(\eta_e = 72.733\) GFlops/W, where this unit denotes \(10^9\) floating-point operations per second per watt. For the quantum hardware, we consider an air-cooled cryostat system, specifically the AttocubeCMC \footnote{\url{https://www.attocube.com/en/products/cryostats/compact-mobile-cryogenics/attoCMC}}, which has a cooling power $P_0=50$ mW at $2.8$\,K and an efficiency $\eta_c$ equal to $x=0.4\%$ of the Carnot efficiency. From Eq.~\eqref{eq:p-cryo}, this means that a single cryostat requires 1,396 kW to operate at 3\,K \footnote{This includes the power for the cryostat chamber and the compressor.}. This choice is motivated by the fact that small air-cooled cryostats are currently best fitted for photonic quantum computers with single-photon sources. We furthermore assume $s_\text{max}=26$, but this value could be significantly higher for larger cryostats, or using waveguide integrated SNSPDs, which are more compact. The choice $s_\text{max}=26$ implies that $\bar n \leq 12$, meaning that a 1-to-12 active demultiplexer (DMX-12) is sufficient to demultiplex the photons. Finally, we assume that $P_\text{fix}=1$\,kW (in accordance with current experimental values) and that $r_{\text{SPS}}=1$ GHz. 

We are now in a position to compare the energy cost per sample for the classical and quantum cases. We begin by assuming a fixed number of lost photons. A preliminary analysis reveals $l=12$ as a good working hypothesis, being both a reasonably low yet relevant value. We then compute the classical and quantum energy per sample as a function of the target metric value $M_0$ for $l=12$ (the plots are shown in the Appendix~\ref{app:fixednumber}).
The energy per sample for the classical case is computed using Eq.~\eqref{eq:classical-energy-per-sample-2}, setting the metric at $M_0$. The quantum energy per sample is obtained using, in Eq.~\eqref{eq:quantum-energy-per-sample}, the values of the control parameters $(T, n)$ obtained by minimizing the resource $P_{\text{tot}}$ (see Eq.~\eqref{eq:p-tot}) given a fixed metric $M_0$. Calling $n^*$ the optimal result for $n$, this procedure gives a requirement on the end-to-end transmission of the setup, $\eta^*\equiv (n^*-l)/n^*$. The range of the parameters are $1\text{K}<T_{\text{cryo}}<3 \text{K}$ -- ensuring that the source and detectors operate -- and $l+1\leq n\leq 200$. We find that $E_\text{sample}^\text{C}$ scales exponentially with $M_0$ (as expected), while $E_\text{sample}^\text{Q}$ exhibits a more favorable scaling. 

In the discussion above, we assumed that $l=12$ photons are lost. However, in practice, the number of transmitted photons in a setup of transmission $\eta^*$ fluctuates. This, in turn, induces fluctuations of the metric $\mathcal{M}$.  
To account for this effect, we do the following: for each target value $M_0$,  
we compute the total classical energy required to produce a set of $N=10,000$ samples where the number of transmitted photons fluctuates following a binomial distribution of mean $\eta^* n^*$ and standard deviation $\sqrt{n^*\eta^*(1-\eta^*)}$. We normalize this energy by $N$, and compare it to the quantum energy per sample (the quantum energy per sample does not vary with the number of lost photons). The results are shown on Fig.~\ref{fig:results}.(a), where the quantum and average classical energies per sample are plotted as functions of the average metric. The green line at $M_e^*=15$ marks the energy advantage threshold: for higher values of the average metric, the quantum computer consumes less energy than the classical one.

We highlight that the advantageous scaling of $E_\text{sample}^\text{Q}$ comes at the expense of requiring increasingly high transmission as the metric increases. In other words, increasing the metric requires progressively larger photonic chips with ever-improving performance. In practice, however, the transmission decreases exponentially with the number of optical components, with a rate depending on the attenuation per component (see the following section and Eq.~\eqref{eq:eff}). 
We will examine the experimental feasibility in the following section, but let us now give a qualitative discussion on how the metric $\mathcal{M}$ and energy cost are expected to evolve with the size of the system (i.e.~number of modes) at a fixed transmission value. Let us assume that we make a series of experiments with systems of increasing size, specifically with $m=\lceil 2.1 n\rceil$ modes for $n\in [1,75]$, but with a constant attenuation per component. 
The results are shown on Fig.~\ref{fig:results}.(c),(d): if we do not increase the end-to-end transmission as we increase the system size, the metric drops for larger values of input photons since loss becomes overwhelming.

In order to compare with a computational advantage, we now turn to the time performance of the quantum and classical computers.   
The time required by the classical computer to produce a sample is given by
\begin{equation}
    t^C_{\text{sample}}(T,n) = \frac{L[\lceil 2.1 n\rceil,\mathcal{M}(T,n)]}{R_{\text{max}}}
\end{equation}
where $R_{\text{max}}$ is the number of PFLOP per second. For the JEDI module, we have $R_{\text{max}} = 4.5$ PFLOP/s. In the quantum case, the time per sample is simply the inverse of the sample generation rate given in Eq.~\eqref{eq:q-rate},
\begin{equation}
    t_{\text{sample}}^Q(n)=\frac{1}{r_\text{sample}(n)} \, .
\end{equation}
We follow the same procedure as for the energy, studying first the case of $l=12$, then accounting for the fluctuations to compute the average classical time per sample.
The results are shown on Fig.~\ref{fig:results}.(b). The red line at $M^*_c = 18$ represents the threshold for the regime of quantum computational advantage: beyond this point, quantum boson samplers generate samples faster than the best-known classical algorithms \footnote{The computational advantage threshold was predicted to occur at around $M_0=50$ \cite{clifford2018classical, wu2018benchmark}. However, in those works, the prediction was for Boson Sampling in the no collision regime ($m=n^2$, as opposed to $m=\lceil 2.1 n\rceil$ here) and in \cite{wu2018benchmark} the estimation was based on the performance of supercomputers of 2016, but the efficiency of these machines has much evolved since.}. Remarkably, our analysis reveals the existence of a genuine quantum energy advantage for $15\leq M_0\leq 18$ (gray band in Fig.~\ref{fig:results}.(a),(b)): in this window, the quantum computer produces a sample slower than the classical computer but uses less energy.

\medskip

%% file: sections/4-Hardware.tex
\subsection{Experimental feasibility}
We now turn to the experimental feasibility of reaching the energy and computational advantage. As shown in Fig.~\ref{fig:results}.(a) and Fig.~\ref{fig:results}.(b), accessing the quantum advantage regimes (energetic or computational) requires simultaneously increasing both the system size and the transmission efficiency. In practice, however, transmission inevitably decreases as the optical circuit grows.  
To see this, let us derive a realistic expression for the transmission efficiency as a function of the system size.
Photon loss arises mainly from two sources: (i) coupling losses at the fiber–waveguide interfaces and (ii) propagation losses within the circuit.
The coupling loss per interface, $c_\text{coup}$ in dB, depends on the fabrication technology and materials used. 
The propagation loss is determined by the optical depth of the chip, $d(m)$,
which in turn depends on the number of components the photons go through (which scales with the number of modes), and on the loss per Mach-Zehnder interferometer (MZI) on the chip, $c_\text{MZI}$ dB/MZI. 
The optical depth $d(m)$ depends on the architecture of the chip. 
For the commonly used Clements architecture, the optical depth is $d(m)=m$.
In addition to coupling and propagation loss, the overall transmission efficiency also depends on the brightness of the source at the output fiber ($\eta_\text{of}$), on the efficiency of the SNSPDs ($\eta_\text{d}$), and on the efficiency of the demultiplexer ($\eta_\text{dmx}$). Altogether, the transmission is given by 
\begin{equation}
    \eta(m) = \eta_\text{of}\eta_\text{d}\eta_\text{dmx} 10^{-2c_\text{coup}/10} 10^{-mc_\text{MZI}/10} \, .
    \label{eq:eff}
\end{equation}
There is a factor of 2 before $c_\text{coup}$ because there are two coupling interfaces (entry and exit of the chip).

In the previous section, we found that the energy advantage requires producing samples with an average metric $M_e^*=15$.
This can be achieved using the setup depicted on Fig.~\ref{fig:circuit} with an end-to-end transmission of 60\%, 51 modes and two sources placed in two cryostats, and using 24 input photons. 
As shown in Table.~\ref{table:source-eff-target}, current state-of-the-art devices are not yet capable of reaching the quantum energy advantage threshold. However, modest improvements, in particular higher-efficiency demultiplexers, would be enough to bridge the gap: see the `Energetic adv.' column of Table.~\ref{table:source-eff-target}. For the computational advantage, we require two cryostats with 24 detectors each and one cryostat with 11 detectors (59 detectors in total), 3 DMX-12, and sending 28 photons of indistinguishability $I=95\%$ in a setup of end-to-end efficiency 65\% (see Appendix~\ref{app:expfeas} for details). These requirements can be achieved in the near future, as shown on the last column of the Table.~\ref{table:source-eff-target}. We mention that there are several ways to achieve the target transmissions in Table.~\ref{table:source-eff-target}; we chose, for the last two columns, to use near-term efficiencies for demultiplexers and sources, in order to point out that this would allow us to relax the requirement on the transmission per optical component ($c_\text{MZI}$).

\begin{table}[t]
\centering
\resizebox{\linewidth}{!}{%
\begin{tabular}{|c|c|c|c|c|c|}
\hline  
  &   State of the art & Energetic adv. & Comp. adv.\\ 
  \hline
$\eta_\text{of}$   & 71.2\%  \cite{ding2025aboveloss} &  75 \% & 80\% \\
\hline
$\eta_\text{d}$ & 98\% \cite{zadeh2021snspd} &  99 \% & 99\% \\
\hline
$\eta_\text{dmx}$ & 83\% \cite{hardwareRequirements}  & 92\% \cite{uppu2020sclalableintegrated}& 95\% \\
\hline
$c_\text{MZI}$ & 0.0035 dB \cite{wang2019bs-20-photons} & 0.0085 dB & 0.009 dB \\
\hline
$c_\text{coup}$  & 0.057 dB \cite{psiq2025platform} &  0.06 dB & 0.057 dB \\
\hline \hline
$\eta(m)$ & 54\%  &  60\% & 65\% \\
\hline
$I$  & $93\pm 0.8\%$ \cite{zhai2022interferenceremoteqd} &  95\% & 95\% \\
\hline
\end{tabular}%
}
\caption{Detailed noise budget with state-of-the-art numbers, and modest improvements required to reach the energetic (middle column) and computational (right column) advantages (see Eq.~\eqref{eq:eff}). From left to right, $\eta$ is computed for $m=51, 51$ (required for an energetic advantage demonstration), and $m=59$ (for computational advantage demonstration).
The last line corresponds to indistinguishability of single photons $I$.}
\label{table:source-eff-target}
\end{table}

Far from being a distant or purely theoretical milestone, the quantum energy advantage therefore appears as an experimentally attainable goal in the near future.

%% file: sections/5-sycamore.tex
\subsection{Energetic comparison between photonic and superconducting modalities}
Finally, we comment on the energetic performance of photonic quantum computers versus superconducting quantum computers for sampling tasks. Boson Sampling is a sampling task suited to photonic quantum computers. A counterpart for superconducting quantum computers is the Random Circuit Sampling (RCS) problem \cite{boixo2018supremacy, bouland2019rcs}: starting from a set of qubits prepared in a given state, apply a random sequence of one- and two-qubit gates, and measure the qubits in the computational basis. 
Computing the probability distribution of the outputs classically becomes exponentially hard with the number of qubits and gates, but producing the samples can be done natively with gate-based quantum computers.

Any comparison between Boson Sampling and Random Circuit Sampling must be interpreted with care. These tasks rely on distinct physical models, noise mechanisms, and complexity assumptions, and to our knowledge no formal mapping between their hardness metrics is currently known. The comparison presented here is therefore not intended as a strict equivalence, but rather as an order-of-magnitude, hardware-level benchmark of the energetic cost required by different quantum platforms to generate samples in representative quantum advantage experiments. Since both are paradigmatic quantum advantage demonstrations, a natural common ground for benchmarking is the absolute runtime of the best-known classical algorithm for producing a sample. While such a metric inevitably depends on the chosen classical strategy, this dependence is intrinsic to quantum advantage demonstrations.

Interestingly, the Sycamore experiment \cite{arute2019rcs} reports that the energetic cost per sample for RCS is essentially constant, independent of the number of qubits or circuit depth, because all samples are generated within a single cryostat and the dynamic heat load is negligible. Consequently, even without an explicit mapping between Boson Sampling and RCS metrics, the energy cost for superconducting implementations can be treated as constant, simplifying this cross-platform energetic comparison.
On the Sycamore processor, the energy cost required to generate $10^6$ samples was evaluated at 1\,kWh, and it took 200\,s (see Supplemental Material of \cite{arute2019rcs}); hence $10^{-3}$\,Wh (and $2\cdot 10^{-4}$s per sample) which is \emph{five orders of magnitude} more than the photonic platform, see Fig.\ref{fig:results}.(a). 
We want to recall that these Sycamore values, obtained from an actual experiment rather than a projection, provide a useful point of reference but were not optimized for energy performance.
This cross-platform analysis illustrates both the feasibility and importance of benchmarking quantum technologies against one another, not only for computational speed but also for their energy footprint.

%% file: sections/6-conclusion.tex
\section{Conclusions}

We have presented a quantitative analysis of the energetic efficiency of a photonic quantum computer, using Boson Sampling as a case study. By developing a full-stack resource model that incorporates cryogenic power, laser power, and control electronics consumption, and by applying the Metric–Noise–Resource methodology, we have established a direct connection between experimental control parameters, noise processes, and energy consumption for a concrete quantum computational task.

Our analysis demonstrates the existence of a quantum energetic advantage, showing that the energy cost per sample of a photonic Boson Sampling experiment can scale more favorably than that of classical simulation algorithms. Remarkably, this energetic advantage emerges in a regime where photonic quantum computers do not yet exhibit a computational advantage in runtime, highlighting energetic efficiency as an independent and complementary notion of quantum advantage.

We have deliberately adopted a conservative approach in estimating classical energy costs, by considering the most energy-efficient supercomputing hardware available, using a lower bound on the number of floating-point operations, and relying on one of the fastest known classical simulation algorithms. Under these assumptions, classical energy consumption is likely underestimated. By contrast, the energetic efficiency of photonic quantum computers can be further improved through architectural choices such as hosting multiple sources per cryostat or adopting more efficient cooling strategies at scale.

Beyond identifying an energetic advantage in principle, we have translated our analysis into experimentally relevant requirements by providing a detailed noise and loss budget for a near-term Boson Sampling architecture. This shows that observing a quantum energetic advantage is not a distant or purely theoretical goal, but an experimentally attainable milestone with current or near-future photonic technology.

More broadly, our results suggest that energetic efficiency provides a meaningful and experimentally accessible benchmark for quantum advantage, alongside —- but not implied by —- computational speedups. From this perspective, photonic quantum computing platforms appear particularly promising for energy-efficient quantum information processing, and energetic considerations may play a central role in guiding the development and evaluation of future quantum technologies.

\vspace{1cm}
\paragraph*{Acknowledgments.---} This work has been co-financed by the OECQ i-démo project as part of France 2030. 
This research aligns with the QEI's \footnote{\href{https://quantum-energy-initiative.org}{https://quantum-energy-initiative.org}} mission to establish standardized, transparent methodologies for assessing and improving the energy efficiency of quantum technologies. 
We would like to warmly thank our OECQ partners Alexia Auffèves, Robert Whitney, Joseph Mikael, and Jeremy Stevens.
We also warmly thank Olivier Ezratty, Edouard Ivanov, Simone Piacentini, Andreas Fyrillas, Sébastien Boissier, and Rawad Mezher for useful discussions. 

Some of the methods described in this paper are the subject of a pending patent application.

%% file: sections/7-appendix.tex
\section{Appendix}
\label{app}

\subsection{Distinguishability}
\label{app:dist}

We derive here the expressions of $\gamma$ and $\gamma^*$ in Eq.~\eqref{eq:ZPL}, and provide some details on the multiple-source case.

\begin{figure}
    \centering
    \includegraphics[width=1\linewidth]{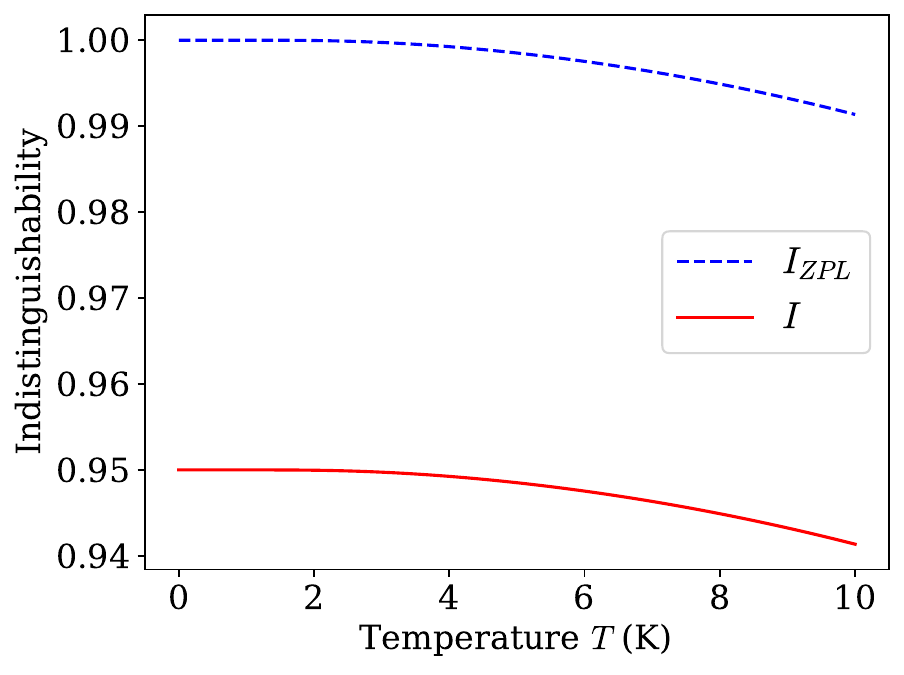}
    \caption{Relationship between indistinguishability and temperature. The blue dashed line represents the indistinguishability of the ZPL, given in Eq.\eqref{eq:ZPL}, while the red line shows the effective indistinguishability considered in our model, given in Eq.\eqref{eq:ZPL-2qd}, accounting for additional noise between distant sources.}
    \label{fig:indist}
\end{figure}

The broadening of the ZPL, i.e. an increase of $\gamma^*$, is due to the interaction with thermal acoustic phonons, which induce a transition to higher energy levels. The scattering of a thermal acoustic phonon to another mode during this process implies a temperature dependence of the form 
\begin{align}
    \gamma^*(T) = \alpha n_{\text{BE}}(T) \left(n_{\text{BE}}(T) + 1\right) \, ,
\end{align}
with \( n_{\text{BE}}(T) = 1 / (\exp\left(\frac{\varepsilon_p}{k_B T}\right) - 1 ) \) the Bose-Einstein distribution, and $k_B$ the Boltzmann constant. $\varepsilon_p=1$\,meV is the energy associated with the wavelength of maximally coupled acoustic phonons, and $\alpha$ is an empirical rate describing the scattering with thermal acoustic phonons. We consider $\alpha=0.1\,\mu$eV in accordance with \cite{grange2017reducing}. The rate $\gamma(T)$ in Eq.~\eqref{eq:ZPL} is given by
\begin{align}
    \gamma(T) = [1 + \eta_{\text{ZPL}}(T) \cdot F_{\text{eff}}(T)] \gamma_0 \, ,
\end{align}
with \( \gamma_0 = 0.658 \, \mu\)eV the natural linewidth of the QD without a cavity, \(\eta_{\text{ZPL}}\) the emission fraction into the ZPL in the presence of a cavity (which decreases linearly with temperature \cite{grange2017reducing}) and \(F_{\text{eff}}\) denotes the effective Purcell factor, modeled under the assumptions that the quantum dot is kept in resonance with the cavity and that phonon sidebands (PSBs) are absent \cite{grange2017reducing, grange2017reducingsupp}. This leads to
\begin{equation}
\gamma_0 F_{\text{eff}}(T) = \frac{4g^2}{\kappa+\gamma_0+\gamma^*(T)} \, .
\end{equation}
In this equation, \(\kappa = 90 \, \mu\)eV is the cavity linewidth and \(g = 90\) \slash \(\sqrt{2} \, \mu\)eV is the QD-cavity coupling strength. The indistinguishability of the ZPL is plotted on Fig.\,\ref{fig:indist}.

Let us now consider the indistinguishability between two photons emitted by two distant sources, labeled $A,B$. Setting aside the slow frequency fluctuations for now, the indistinguishability is upper bounded by the minimum between $\sqrt{I_\text{ZPL}^A I_\text{ZPL}^B}$, and the classical wavepacket overlap, $4\gamma_A\gamma_B/(\gamma_A+\gamma_B)^2$ (where $I_{A,B}$ and $\gamma_{A,B}$ are respectively the ZPL indistinguishability and the decay rates of the sources $A,B$) \cite{pont2025remoteqd}. In this work, we assume that the sources are nearly identical and that the noise due to pure dephasing dominates, meaning that $\sqrt{I_\text{ZPL}^A I_\text{ZPL}^B}\approx I_{ZPL}$ and $\sqrt{I_\text{ZPL}^A I_\text{ZPL}^B}\leq 4\gamma_A\gamma_B/(\gamma_A+\gamma_B)^2$. The slow spectral fluctuations are finally accounted for by removing a fixed value 5\% to the indistinguishability, yielding the expression in Eq.~\eqref{eq:ZPL-2qd}, and plotted on Fig.\,\ref{fig:indist}.

\subsection{Analysis at fixed number of lost photons}
\label{app:fixednumber}

\begin{figure}[h]
    \centering
    \includegraphics[width=1\linewidth]{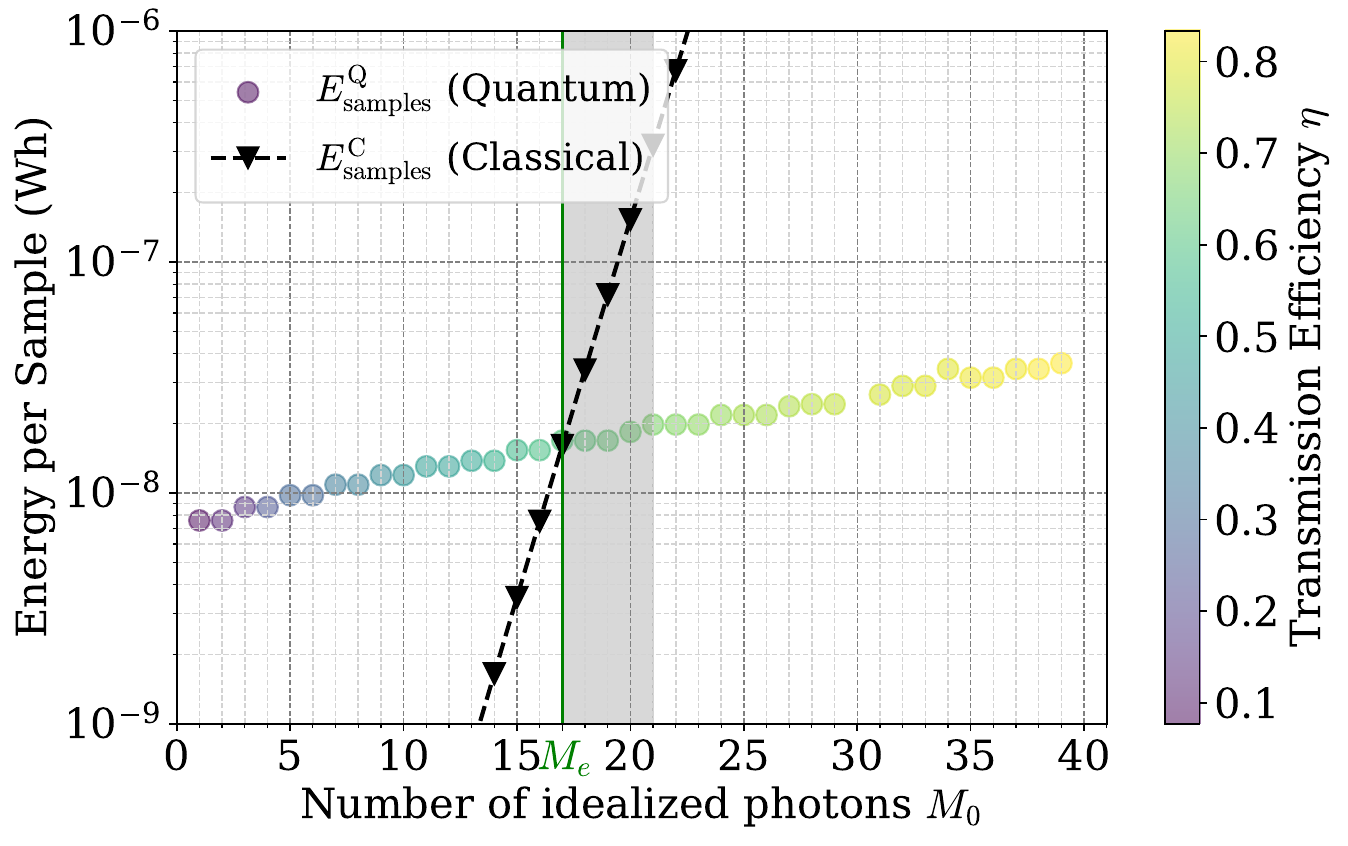}
    \includegraphics[width=1\linewidth]{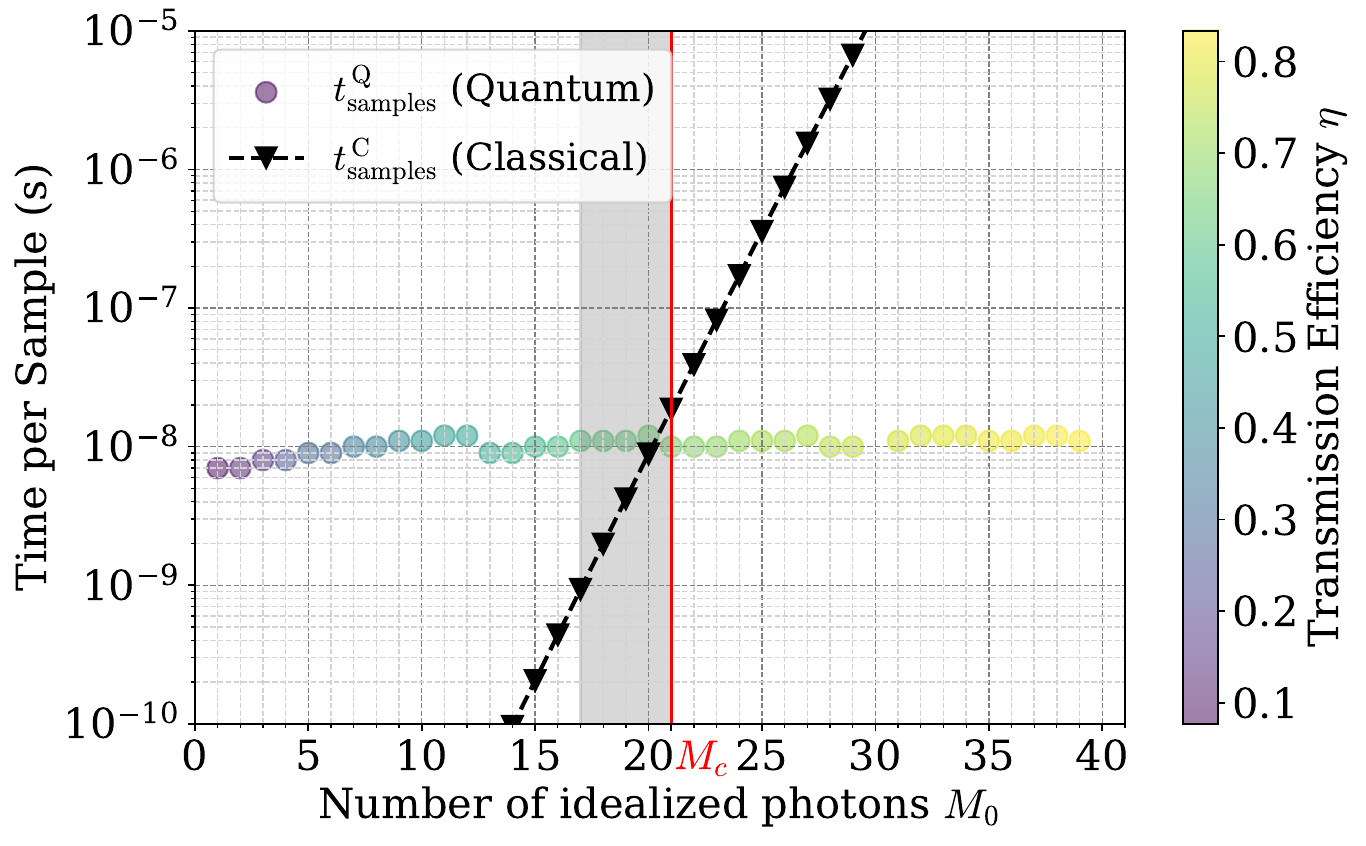}
    \caption{Top: Energy per sample for quantum ($E^\text{Q}_\text{samples}$) and classical ($E^\text{C}_\text{samples}$) computers as a function of the number of idealized photons $M_0$. The quantum energy per sample points are colored according to the transmission efficiency, as
indicated by the color bar. Bottom: Time per sample for the quantum ($t^\text{Q}_\text{samples}$) and classical ($t^\text{C}_\text{samples}$) computers. The gray
area shows the regime where the photonic quantum computer is slower but more energy efficient than the classical one. }
    \label{fig:res-l-12}
\end{figure}

As explained in the main text, we start our analysis by assuming a fixed number of lost photons, and then use the value $l=12$ to compute the required transmission efficiencies. The motivation is the following: increasing $l$ allows to span a wider range of values for the transmission efficiency, $(n-l)/n$, which comes into play in the metric. 
Thus, we can perform a more refined optimization. 
In addition, we would ideally want to have the condition $k^\text{eff}\leq n-l$ satisfied over the range of $n$ used for the optimization. 
Otherwise, the interplay between $n$ and $T$ cannot take place. We therefore tested, for values of $l\in[1,25]$, which values of $k^\text{eff}$ could be reached and on which intervals for $n$. Then, we performed the optimization for each value of $l\in[1,25]$. We find that, (i) the energy advantage threshold is always at $M_0=17$, and (ii) that 
for $l<12$, $k^\text{eff}$ is always strictly larger than $17$. This is the reason why we choose $l=12$ in the main text, since it allows us to reach the energy advantage threshold while genuinely performing an optimization. The lowest value for $k^\text{eff}$ when $l=12$ is 14. We extend the plots to $M_0\in[1,13]$ for better visibility; on the range $M_0\in[1,13]$, however, we highlight that the metric is equal to $n-l$. The results are shown on Fig.~\ref{fig:res-l-12}.

\subsection{Experimental feasibility}
\label{app:expfeas}

We provide details on the methodology used to determine the experimental feasibility. 

\begin{figure}[h]
    \centering
    \includegraphics[width=\linewidth]{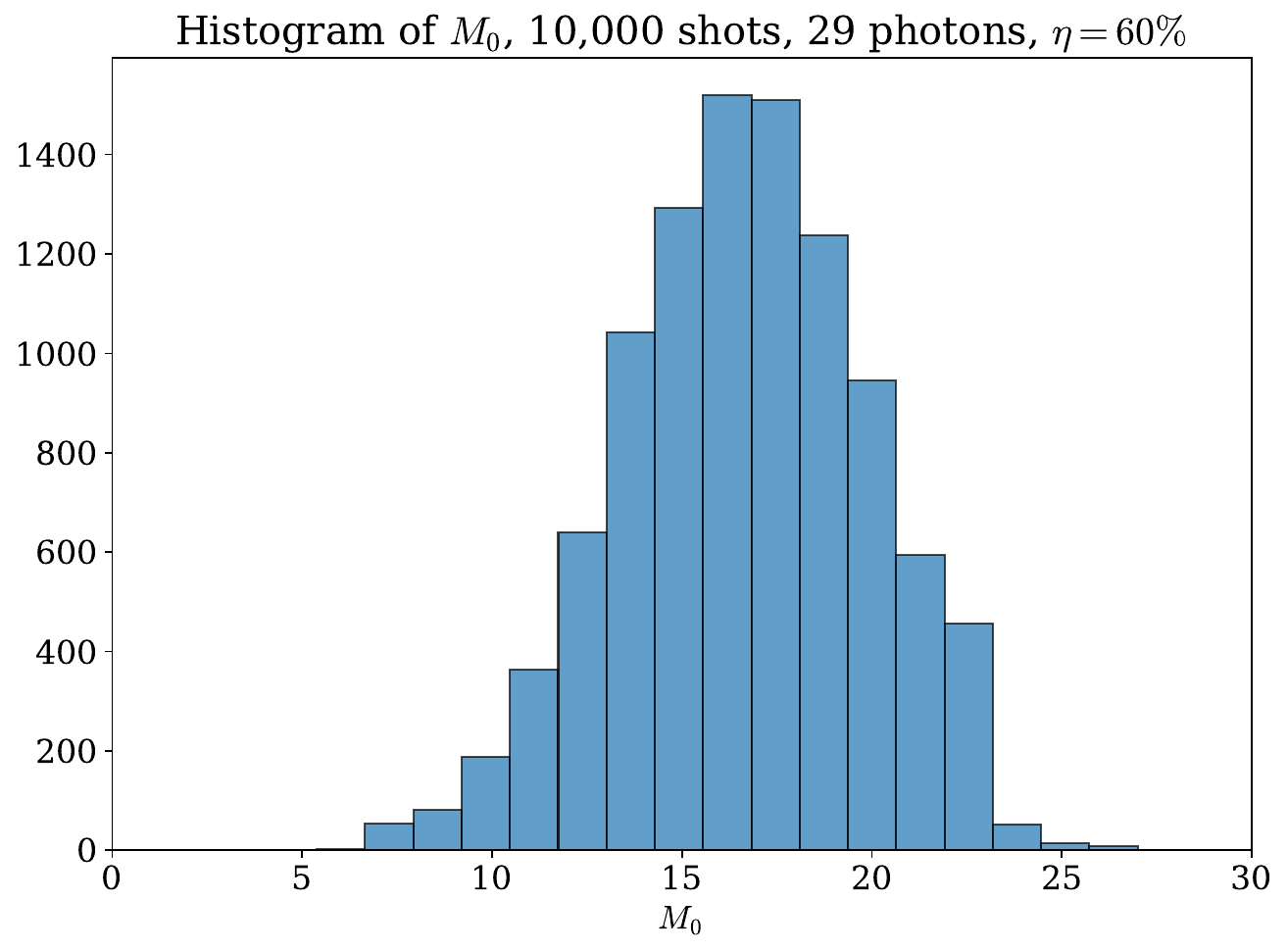}
    \caption{Histogram of the metrics obtained from 10,000 samples with an experimental setup of end-to-end transmission 60\%, 61 modes, and 29 input photons. }
    \includegraphics[width=\linewidth]{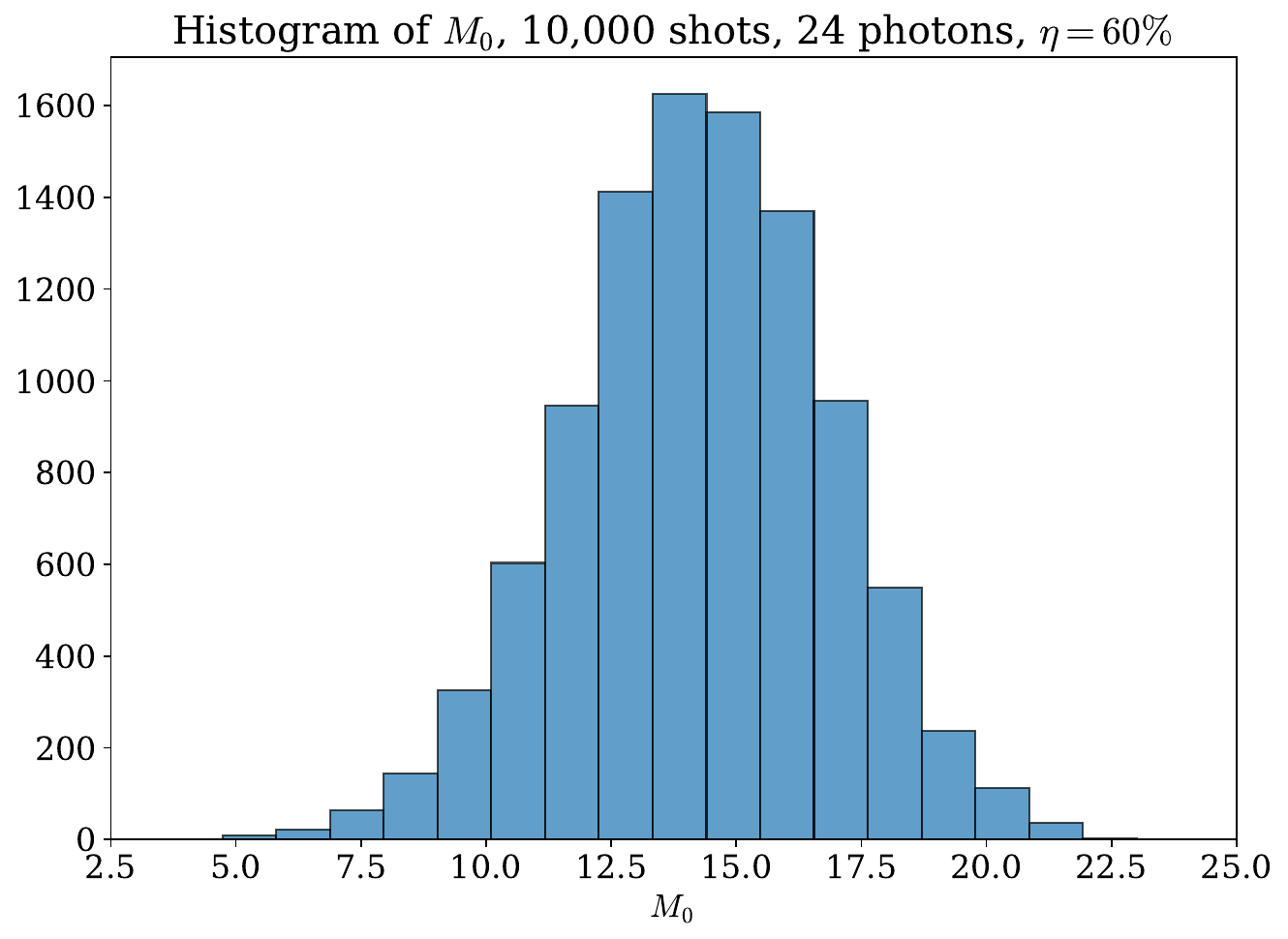}
    \caption{Histogram of the metrics obtained from 10,000 samples with the experimental setup depicted in Fig.~\ref{fig:circuit}, of end-to-end transmission 60\%, 51 modes, and 24 input photons. }
    \label{fig:histogram}
\end{figure}

Let us call $M_e\equiv 17$ and $M_c\equiv 21$ the metrics for the energy and computational advantage thresholds, obtained during the optimization at fixed number $l=12$ of lost photons, see Fig.~\ref{fig:res-l-12}. The optimization revealed that reaching $M_e$ requires an efficiency of 60\%, while reaching $M_c$ requires 65\%. In both cases, the optimal cryostat temperature was found to be $3$\,K, yielding an indistinguishability $I=95\%$. Since we assumed that $l=12$ photons are lost, this means using 29 and 34 photons, respectively. Using a setup of end-to-end transmission $\eta=60\%$ (resp. 65\%), with 61 modes (resp. 72 modes), and sending $n=29$ (resp. 34) photons, would therefore allow to create samples of metric $M_e$ (resp. $M_c$). 

In practice, the number of transmitted photons fluctuate, following a binomial distribution of mean $\eta n$ and standard deviation $\sqrt{n\eta(1-\eta)}$, which in turn induces fluctuations of the metric $\mathcal{M}$, with a standard deviation $\sigma$. 
One could then argue that, in order to experimentally demonstrate an energetic advantage, we would like a certain percentage of the samples produced to have a metric higher than $M_{e,c}$. For instance, if we wanted $68\%$ of the samples with that property, then we would require a setup which produces samples of average metric $M_{e,c}^*=M_{e,c}+\sigma/2$. 
However, since the classical cost per sample grows exponentially with the values of the metric, it is more relevant to compare the quantum energy cost per sample to the average classical cost per sample, $\langle E_\text{samples}^C\rangle$, if one were to classically simulate the distribution of samples produced by the quantum computer. The histogram on Fig.~\ref{fig:histogram} shows the distribution of the metrics obtained in an experiment with a setup of end-to-end transmission $\eta=60\%$, 61 modes, and using $n=29$ input photons. As expected, the metric is centered around $M_e=17$, but we find that the energy consumed by the classical computer is on average per sample $1.3\times 10^{-7}$\,Wh, one order of magnitude larger than the quantum energy per sample, which is $1.4\times 10^{-8}$\,Wh.

We can perform the same averaging for each point in the Fig.~\ref{fig:results}.(a),(b), and we obtain a new energy threshold at $M_e^*=15$, see Fig.~\ref{fig:results}. 

We therefore propose the setup, depicted on Fig.~\ref{fig:circuit}, for an experimental demonstration at the energetic advantage threshold, that is, an experiment where the average classical and quantum energy per sample are equal. To reach the point $M_e^*=15$ on Fig.~\ref{fig:results}.(a) requires 28 input photons and an end-to-end transmission of 56\%. Experimentally, since we have access to DMX-12, it would be more favorable to use 24 photons, as we then only need two sources. Reducing the number of input photons while maintaining an average metric of $M_e^*=15$ and keeping $\langle E_\text{samples}^C\rangle =  E_\text{samples}^Q$ requires increasing the end-to-end transmission to 60\%. Finally, using a setup of end-to-end transmission $\eta=60\%$, 51 modes, and using $n=24$ input photons, we obtain $\langle E_\text{samples}^C\rangle =  E_\text{samples}^Q = 1.0\times 10^{-8}$\,Wh.